\documentclass[prl,aps,10pt,superscriptaddress,twocolumn,floatfix]{revtex4-1}

\usepackage[nice]{nicefrac}
\usepackage{graphicx,color}
\usepackage{amsmath,amssymb,bm,bbm}
\usepackage{amsmath}
\usepackage{braket}
\usepackage{bbold}
\usepackage{float}

\usepackage[plainpages=false,pdfpagelabels,colorlinks=true,linkcolor=red,urlcolor=blue,citecolor=blue,pdftitle={},pdfauthor={},pdfdisplaydoctitle=true,pdfduplex=DuplexFlipLongEdge]{hyperref}

\definecolor{darkred}{rgb}{0.90,0.2,0.2}
\definecolor{darkgreen}{rgb}{0,0.60,.2}
\definecolor{darkblue}{rgb}{0.1,0.3,1}
\definecolor{grey}{cmyk}{0,0,0,0.25}
\definecolor{orange}{cmyk}{0,0.6,0.8,0}

\begin{document}

\title{Scale-Invariant Survival Probability at Eigenstate Transitions}

\author{Miroslav Hopjan}
\affiliation{Department of Theoretical Physics, J. Stefan Institute, SI-1000 Ljubljana, Slovenia}
\author{Lev Vidmar}
\affiliation{Department of Theoretical Physics, J. Stefan Institute, SI-1000 Ljubljana, Slovenia}
\affiliation{Department of Physics, Faculty of Mathematics and Physics, University of Ljubljana, SI-1000 Ljubljana, Slovenia\looseness=-1}

\begin{abstract}
Understanding quantum phase transitions in highly excited Hamiltonian eigenstates is currently far from being complete.
It is particularly important to establish tools for their characterization in time domain.
Here we argue that a scaled survival probability, where time is measured in units of a typical Heisenberg time, exhibits a {scale-invariant behavior} at eigenstate transitions.
We first demonstrate this property in two paradigmatic quadratic models, the one-dimensional Aubry-Andre model and three-dimensional Anderson model.
Surprisingly, we then show that similar phenomenology emerges in the interacting avalanche model of ergodicity breaking phase transitions.
This establishes an intriguing similarity between localization transition in quadratic systems and ergodicity breaking phase transition in interacting systems.
\end{abstract}

\maketitle

\textit{Introduction.--}
Quantum phase transition in highly excited Hamiltonian eigenstates (henceforth, eigenstate transitions) can be seen as a generalization of ground-state quantum phase transitions~\cite{sachdevbook}.
They are often characterized by an abrupt change of certain wave function properties such as participation ratios or entanglement entropies.
Some remarkable consequences of eigenstate transitions may be manifested in nonequilibrium quantum dynamics of isolated~\cite{polkovnikov11, Eisert2015} or Floquet~\cite{Hedge14,Das17,Haldar17,Haldar18,Haldar21,Haldar22} quantum systems, and may call for refinement of our understanding of quantum chaos~\cite{haake_gnutzmann_18, stockmann2000quantum, dalessio_kafri_16} and thermalization~\cite{dalessio_kafri_16, mori_ikeda_18, deutsch_18}.

In time domain, the overlap of two time-evolving quantum states may represent a useful probe to study the properties of Hamiltonians that govern the dynamics.
Generally, stability of isolated quantum systems against perturbations is studied within the concept of fidelity or Loschmidt echo \cite{Peres84}, which became one of the most important
tools in the theory of quantum chaos \cite{Gorin_06,Goussev_12} and other areas of physics~\cite{Goussev_12}.
Here, we focus on survival probability~\cite{Ketzmerick_92}, which is the squared overlap of the time-evolving state with its initial state, whose main features (e.g., the slopes of its decay) can also be extracted~\cite{LozanoNegro2021}, for small systems, from experimental protocols based on Loschmidt echoes~\cite{LozanoNegro2021,Sanchez2020}.
Of particular interest are its properties at intermediate and long times, which may carry nontrivial fingerprints of eigenstate transitions~\cite{Ketzmerick_92,Schofield_95,Schofield_96}.

In the context of quadratic Hamiltonians in which eigenstate transitions are driven by disorder, a large amount of previous studies focused on survival probability~\cite{Ketzmerick_92,Schofield_95,Schofield_96,Ketzmerick_97,Gruebele_98,Ng_06,Leitner_15,Karmakar_19}.
Perhaps the most important outcomes of these studies are
(i) emergence of a power-law behavior close to and at the eigenstate transition~\cite{Ketzmerick_92,Schofield_95,Schofield_96,Ketzmerick_97,Gruebele_98,Ng_06,Leitner_15,Karmakar_19},
and (ii) connecting the power-law exponent to the fractality of the wave function~\cite{Geisel_91,Ketzmerick_92,Huckestein94,delRio_95,Ketzmerick_97,Kravtsov10,Kravtsov11,Kravtsov12}.
It appears that these properties do not crucially depend on whether the quadratic Hamiltonian is local (such as the Anderson and Aubry-Andre models) or it is given by a random-matrix-theory type of model~\cite{Ng_06,Bera18}.
In spite of these activities, however, it remains unclear whether a power-law decay of survival probability is a sufficient criterion for a detection of the transition point.

Survival probability in interacting systems has not yet received as much attention as in quadratic systems, apart from several exceptions~\cite{Torres-Herrera_14,Santos2017,torresherrera_garciagarcia_18, Prelovsek18, Schiulaz_19,Lezama_21}. 
In random-field spin-1/2 Heisenberg chains, emergence of a power-law decay was reported for a broad range of disorder strengths~\cite{Torres-Herrera_14, torresherrera_garciagarcia_18, Prelovsek18, Schiulaz_19,Lezama_21}, suggesting that the power-law survival probability {\it per se} may not be sufficient to pinpoint the transition in finite systems.
However, the quest for exploring the boundaries of thermalization and the emergence of nonergodic phases of matter has recently experienced tremendous scientific interest~\cite{Rahul15, altman15, abanin2019}.
It is then an urgent task to establish tools to detect eigenstate transitions through the lens of quantum dynamics, both for single-particle and many-body states.

In the context of interacting systems, it is currently not obvious which are the prototypical models that exhibit an ergodicity breaking phase transition in the thermodynamic limit and are at the same time not subject to severe finite-size effects in numerical analyses.
One of the most widely studied systems in this respect is the random-field spin-1/2 Heisenberg chain, for which different predictions about the fate of ergodicity breaking phase transition have recently been made~\cite{suntajs_bonca_20a, suntajs_bonca_20b, kieferemmanouilidis_unanyan_20, sels2020, kieferemmanouilidis_unanyan_21, leblond_sels_21, vidmar_krajewski_21, Sels_dilute_2021, krajewski_vidmar_22, Panda2020, Sierant2020, sierant_lewenstein_20, abanin_bardarson_21, corps_molina_21, prakash_pixley_21, schliemann_costa_21, hopjan_orso_21, solorzano_santos_21, detomasi_khaymovich_21, crowley_chandran_22, ghosh_znidaric_22, bolther_kehrein_22, yintai_yufeng_22, sierant2022, Morningstar2022, sutradhar_ghosh_22, trigueros_cheng_22, shi_khemani_22}.
A convenient alternative for such studies can be formulated within the so-called avalanche model of ergodicity breaking phase transitions~\cite{deroeck_huveneers_17, luitz_huveneers_17, thiery_huveneers_18, crowley_chandran_20, Sels_2022, Morningstar2022, suntajs_vidmar_22, crowley_chandran_22b}, which allows for establishing analytical predictions of the value of the transition point~\cite{deroeck_huveneers_17, luitz_huveneers_17}.
Importantly, numerical results in finite systems comply with these predictions and exhibit only mild finite-size effects~\cite{suntajs_vidmar_22}.

In this Letter, by studying quantum dynamics through the perspective of survival probability, we show that its scale-invariant behavior is a hallmark of eigenstate transitions in both quadratic and interacting systems. 
This allows us to establish a connection between eigenstate transitions in disordered quadratic systems and ergodicity breaking phase transitions in interacting systems.
 
Our analysis consists of two steps.
In the first step, we study two paradigmatic quadratic systems, the one-dimensional (1D) Aubry-Andre model and the three-dimensional (3D) Anderson model, and we introduce a scaled survival probability $p(t)$; see Eq.~(\ref{eq:sur_prob_norm}).
With this we benchmark scale invariance of $p(t)$ as an indicator of a disorder-driven localization transition point in quadratic systems.
Then we extend our analysis to an interacting system, i.e., to the avalanche model.
We show that an identically defined $p(t)$, however on many-body wave functions, also exhibits scale invariance at the ergodicity breaking phase transition.
The scale invariance of $p(t)$ allows us to relate the power-law exponent of $p(t)$ to the fractal dimension of initial states in the eigenbasis of Hamiltonian $\hat H$, and the scaling properties of the typical Heisenberg time.
Finally, we also discuss a connection of wave function based dynamical measures of the transition to the spectrum based measures, such as the spectral form factor.

{\it Scaled survival probability.--}
We are interested in quantum quenches from the initial Hamiltonian $\hat H_0$ with eigenstates $\{|m\rangle\}$ to the final Hamiltonian $\hat H$ with eigenstates $\{|\nu\rangle\}$.
The eigenstates correspond to single-particle (many-body) eigenstates in quadratic (interacting) Hamiltonians.
The eigenstate survival probability for a fixed Hamiltonian realization is defined as
\begin{equation}
\label{eq:sur_prob}
P_{m}^{H}(t)=|\langle m | e^{-i\hat Ht} |m \rangle  |^2 =
\bigg| \sum_{\nu=1}^D  |c_{\nu m}|^2  e^{-iE_\nu t}  \bigg|^2,
\end{equation}
where we set $\hbar \equiv 1$, $D$ is the Hilbert-space dimension, $c_{\nu m}=\langle \nu | m \rangle$ is the overlap of $|m\rangle$ with $| \nu \rangle$, and $E_\nu$ is an eigenenergy of $\hat H$.
The averaged survival probability is defined as
$P(t)=\langle\langle P_m^{H}(t)\rangle_m\rangle_{H}$,
where $\langle ... \rangle_m$ denotes the average over {\it all} eigenstates $|m\rangle$ of the initial Hamiltonian $\hat H_0$, and $\langle ... \rangle_H$ denotes the average over different realizations of the final Hamiltonian $\hat H$.

At long times, $P(t)$ approaches the average inverse participation ratio of eigenstates of $\hat H$ in the eigenbasis of $\hat H_0$, $\overline{P_{}^{}}=\langle\langle \sum_\nu  |c_{\nu m}|^4 \rangle_m\rangle_{H}$.
We express $\overline{P}$ as
\begin{equation} \label{def_Pbar}
\overline{P} = P_\infty + c D^{-\gamma} \;,
\end{equation}
i.e., as a sum of the nonzero asymptotic value $P_\infty=\lim_{D \rightarrow \infty} \overline{P_{}^{}}$ and a part that vanishes in the thermodynamic limit $D \to \infty$ as $\propto D^{-\gamma}$, where $\gamma>0$ is the fractal dimension.
In the fully delocalized regime one gets $P_\infty = 0$, while $P_\infty > 0$ in the localized regime or the regime with a mobility edge.
If the initial wave function at the transition exhibits (multi)fractal properties in the eigenbasis of $\hat H$, one expects $\gamma < 1$.

These considerations allow us to define our central quantity, the scaled survival probability $p(t)$, henceforth survival probability,
\begin{equation}
\label{eq:sur_prob_norm}
p(t)= \frac{P(t)-P_\infty}{\overline{P_{}^{}}-P_\infty} \;,
\end{equation}
which saturates at long times to $\lim_{t\to\infty} p(t) = 1$.
We study $p$ in units of scaled time $\tau = t/t_{H}^{\rm typ}$, where $t_{H}^{\rm typ}=2 \pi/\delta E^{\rm typ}$ is the typical Heisenberg time, $\delta E^{\rm typ}= \exp[\langle \langle \ln( E_{\nu+1} -E_\nu) \rangle_{\nu} \rangle_{H}]$ is the typical level spacing, and $\langle ... \rangle_\nu$ denotes the average over all pairs of nearest levels. 

\textit{Models.--}
We study two quadratic models with particle-number conservation that exhibit localization-delocalization transitions, given by the Hamiltonian
\begin{equation}
\label{eq:ham3DA}
\hat H= -{J}\sum_{\langle ij\rangle}^{} (\hat{c}_{i}^{\dagger}\hat{c}_{j}^{}+ \hat c_j^\dagger \hat c_i) + \sum_{i=1}^{D}\epsilon_{i}\hat{n}_{i}^{}\;,
\end{equation}
where $\hat{c}_{j}^{\dagger}$ ($\hat{c}_{j}^{}$) are the fermionic creation (annihilation) operators at site $j$, $J$ is the hopping matrix element between nearest neighbor sites,
$\hat{n}_{i}^{}=\hat{c}_{i}^{\dagger}\hat{c}_{i}^{}$ is the site occupation operator, and $\epsilon_{i}$ is the on-site energy.
The first is the Aubry-Andre model on a 1D lattice with $L$ sites ($D=L$) subject to the quasiperiodic on-site potential $\epsilon_{i}=\lambda \cos(2\pi q i+\phi)$, where $\lambda$ is
the amplitude of the potential, $q=\frac{\sqrt{5}-1}{2}$ is the golden ratio, and $\phi$ is a global phase.
The model exhibits a sharp localization-delocalization transition at $\lambda_c/J = 2$ for all single-particle eigenstates~\cite{Aubry80, Suslov82, Kohmoto83, Chao86, Kohmoto87, Siebesma1987, Hiramoto89, Hiramoto92, Macia2014, Li16, wu2021} as a consequence of self-duality 
at the transition.
This transition was observed experimentally using cold atoms~\cite{Roati08,Luschen18} and photonic lattices~\cite{Lahini:PRL2009}.
The second is the Anderson model on a 3D cubic lattice ($D=L^3$) subjected to independent and
identically distributed on-site energies drawn from a box distribution $\epsilon_i \in [-W/2,W/2]$.
Numerical studies of transport properties of single-particle eigenstates at the center of energy band~\cite{kramer_mackinnon_93, MacKinnon81, MacKinnon83} based 
on the transfer-matrix technique have shown that the system is insulating for $W >W_{c}  \approx 16.54 J$ \cite{Ohtsuki18} and below $W_{c}$ it becomes diffusive \cite{Ohtsuki97,Zhao20,Herbrych21}.
At the transition, the model exhibits subdiffusion \cite{Ohtsuki97} and multifractal single-particle  eigenfunctions~\cite{evers_mirlin_08, Rodriguez09, Rodriguez10}.
The transition point is energy dependent, i.e., at $W >W_{c}$ all single-particle states are localized, while at $W <W_{c}$ the system exhibits a mobility edge~\cite{Schubert_05}.

We complement our analysis by studying an interacting model that exhibits an ergodicity breaking phase transition, i.e., the avalanche model~\cite{deroeck_huveneers_17, luitz_huveneers_17, suntajs_vidmar_22}.
The model consists of $N+L$ spin-1/2 degrees of freedom in a Fock space of dimension $D=2^{N+L}$.
It is divided into a dot with $N$ spins and a remaining subsystem with $L$ spins outside the dot, described by the Hamiltonian
\begin{equation}
\label{eq:hamQSM}
\hat H= \hat R+g_0\sum_{i=0}^{L-1} \alpha^{u_i} \hat{S}_{n_i}^{x}\hat{S}_{i}^{x}+\sum_{i=0}^{L-1}h_{i}\hat{S}_{i}^{z} \;.
\end{equation}
The spins outside the dot are subject to local magnetic fields $h_i \in [0.5,1.5]$ that are drawn from a box distribution.
Interactions within the dot denoted by $\hat R$ are all-to-all and they exclusively act on the dot subspace.
They are represented by a $2^N \times 2^N$ random matrix drawn from the Gaussian orthogonal ensemble
\footnote{Random matrix $R$ is drawn from the Gaussian orthogonal ensemble:
$R=\beta(A+A^T)/2 \in \mathbb{R}^{2^N\times 2^N}$, where $A_{ij}= {\rm norm}(0,1)$ and ${\rm norm}(0,1)$ are random numbers drawn from a normal
distribution with zero mean and unit variance. We set $\beta=0.3$ as in Refs. \cite{luitz_huveneers_17, suntajs_vidmar_22}. The resulting matrix is embedded to matrix of the full dimension $2^{N+L}$ by Kronecker product
$\hat{R}=R \otimes \mathcal{I}$ where $\mathcal{I}$ is the identity matrix of dimension $2^L$.}.
Each of the spins outside the dot is coupled to one spin in the dot, and the interaction strength is $\alpha^{u_i}$.
For a chosen spin $i$ outside the dot, we randomly select an in-dot spin $n_i$.
The coupling to the first spin outside the dot ($i=0)$ is set to one since $u_0=0$, while at $i \geq 1$, $u_i \in [i-0.2, i+0.2]$ is drawn from a box distribution.

We set $N=5$ and $g_0=1$ in Eq.~(\ref{eq:hamQSM}), and vary the parameter $\alpha$.
For these parameters, the transition estimated from the gap ratio statistics occurs at $\alpha_c\approx 0.716$~\cite{SM}, which is very close to the analytical prediction $\bar\alpha=1/\sqrt{2}\approx 0.707$~\cite{deroeck_huveneers_17, luitz_huveneers_17}. For $\alpha>\alpha_c$ the model is ergodic and it exhibits the Gaussian orthogonal ensemble level statistics~\cite{luitz_huveneers_17, suntajs_vidmar_22}.
This can be interpreted as a successful avalanche induced by the dot.
For $\alpha<\alpha_c$ there is localization of spins outside a thermal bubble and the Poisson level statistics emerges~\cite{luitz_huveneers_17, suntajs_vidmar_22}.

{\it Initial states.--}
Unless stated otherwise, the initial Hamiltonian for quadratic models is $\hat H_0= \sum_{i}^{}\epsilon_{i} \hat n_{i}^{}$, for which the single-particle eigenstates $\{|m\rangle\}$ in Eq.~(\ref{eq:sur_prob}) are fully localized in the site occupation basis.
The survival probability can then be interpreted as a quantity that describes the spreading of a particle initially localized in the disordered lattice.
For the interacting model the initial Hamiltonian is $\hat H_0= {\rm diag}(\hat H)$, i.e., we quench from  $\{|m\rangle\}$ which are product states of fully localized spins in the whole system.
The survival probability hence tracks the stability of the initially localized spins against the avalanche spreading from the dot.

\begin{figure}[t!]
\centering
\includegraphics[width=\columnwidth]{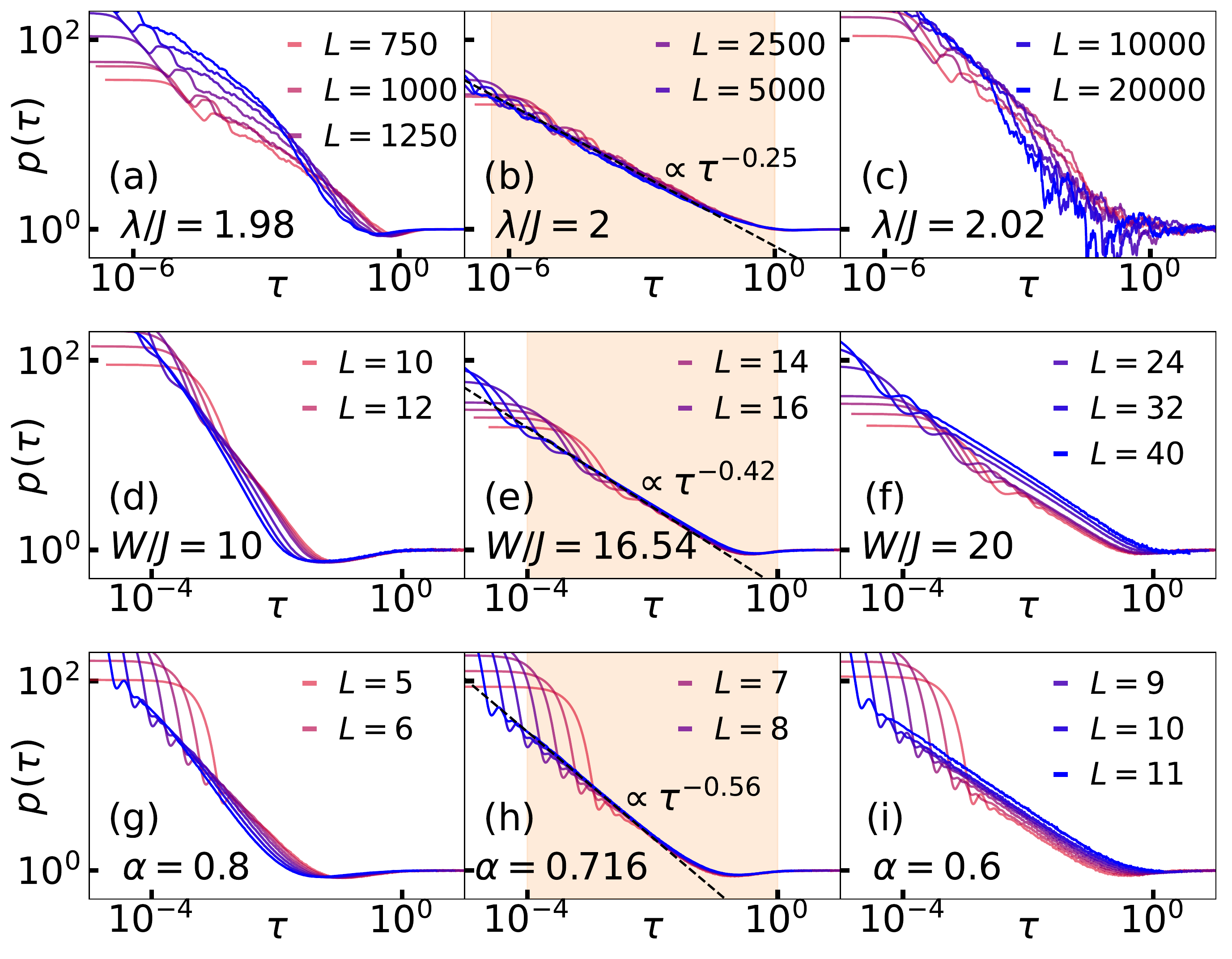}
\caption{
Survival probability $p_{}^{}(\tau)$ as a function of the scaled time $\tau=t/t_{H}^{\rm typ}$ in the 1D Aubry-Andre model (upper panels, a--c),  the 3D Anderson model 
(middle panels, d--f) and the avalanche model (lower panels, g--i) at different system sizes $L$,
 plotted in the delocalized regime (left column), at the transition point (middle column), and in the localized regime (right column).
The shaded areas in the middle column denote the time intervals of the scale-invariant behavior for the largest system sizes.
The dashed lines denote the fits from Eq.~(\ref{def_taubeta}) in the scale-invariant power-law regime.
}
\label{figM1}
\end{figure}

\textit{Scale invariance at the transition.--}
We first study the results in quadratic models.
In the upper panels of Fig.~\ref{figM1} we show $p(\tau)$ in the 1D Aubry-Andre model, while the middle panels of Fig.~\ref{figM1} show $p(\tau)$ in the 3D Anderson model.
At the eigenstate transition, see Figs.~\ref{figM1}(b) and~\ref{figM1}(e), the decay of $p(\tau)$ appears to be independent of the system size.
This scale-invariant behavior extends over several orders of magnitude in time, and it is marked by the shaded areas in Figs.~\ref{figM1}(b) and~\ref{figM1}(e).
We fit the functional dependence in this regime by a power law,
\begin{equation} \label{def_taubeta}
    p(\tau) = a\, \tau^{-\beta} \;,
\end{equation}
where $a$ and $\beta$ are fitting parameters.
We obtain $\beta=0.25$ in Fig.~\ref{figM1}(b) and $\beta=0.42$ in Fig.~\ref{figM1}(e), moreover, in all cases considered in this Letter we obtain $a < 1$.
In contrast, when departing from the transition point toward the delocalized regime, {see Figs.~\ref{figM1}(a) and~\ref{figM1}(d), and toward the localized regime, see Figs.~\ref{figM1}(c) and~\ref{figM1}(f),} scaled-invariant properties are lost and we do not focus on these regimes further on.

We now ask whether a similar behavior can also be observed in an interacting model, i.e., in the avalanche model.
Remarkably, the lower panels of Fig.~\ref{figM1} suggests that this is indeed the case.
Specifically, in Fig.~\ref{figM1}(h) we observe scale-invariant behavior at the ergodicity breaking transition that is fitted by the power law from Eq.~(\ref{def_taubeta}) with $\beta=0.56$. The time interval in which the power law is observed, see the shaded region in Fig.~\ref{figM1}(h), is as broad as that in 3D Anderson model in Fig.~\ref{figM1}(e).
Scale invariance of $p(\tau)$ is lost in the ergodic phase, see Fig.~\ref{figM1}(g), and in the localized phase, see Fig.~\ref{figM1}(i).

\begin{figure}[b!]
\centering
\includegraphics[width=\columnwidth]{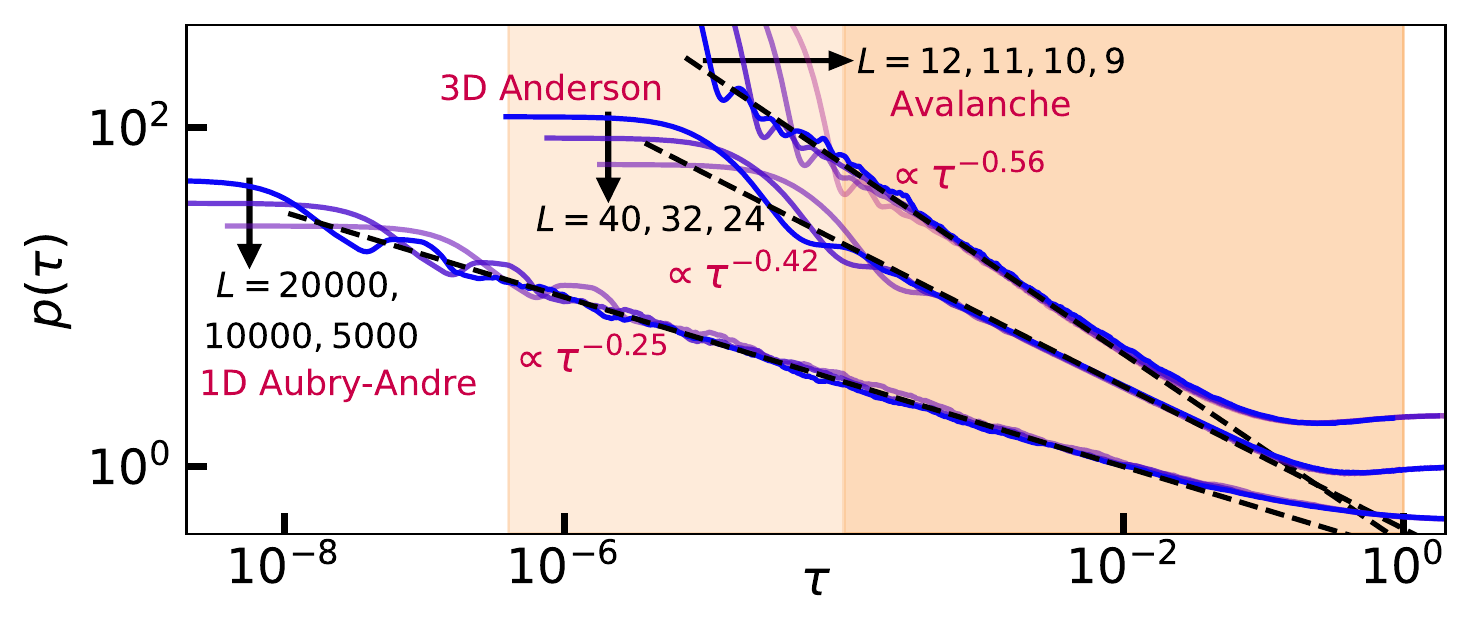}
\caption{
Scale invariance of survival probability $p_{}^{}(\tau)$ at the transition point demonstrated for the largest system sizes $L$ of the models under consideration ({the larger $L$, the darker the color}). 
The curves are identical to those in Figs.~\ref{figM1}(b),~\ref{figM1}(e), and~\ref{figM1}(h), but {shifted in y axis (i.e., multiplied by constants)} for clarity.
The shaded area denotes the time interval of the 
scale invariant behavior for the largest $L$ (for the 3D Anderson model and the avalanche model this time interval roughly coincides).
The dashed lines denote the fits from Eq.~(\ref{def_taubeta}).
}
\label{figM2}
\end{figure}

\textit{Consequences of scale invariance.--}
We now explore the consequences of the observed scale invariance of $p(\tau)$ at eigenstate transitions, which is shown in Fig.~\ref{figM2} for all models under consideration.
We describe the procedure that allows us to relate $\beta$ from Eq.~(\ref{def_taubeta}) to other properties at the transition such as the fractal dimension $\gamma$.

We start by inserting $\overline{P}-P_\infty$ from Eq.~(\ref{def_Pbar}) and the power-law form of $p(t)$ from Eq.~(\ref{def_taubeta}) into Eq.~(\ref{eq:sur_prob_norm}), and considering its logarithm, one obtains $\ln[P(t)-P_\infty] = -\beta \ln t + \ln(a (t_{H}^{\rm typ})^\beta c D^{-\gamma})$.
We note that if the power-law decay of $P(t)-P_\infty$ was to extend until $t=t_{H}^{\rm typ}$ [cf.~the dashed lines in Figs.~\ref{figM1}(b),~\ref{figM1}(e) and~\ref{figM1}(g)], the value $P(t_{\rm H}^{\rm typ}) - P_\infty$ would be lower than $c D^{-\gamma}$ since $a<1$.
However, our goal is to understand the behavior of $\beta$ that corresponds to the slope of the function $\ln[P(t)-P_\infty]$ versus $\ln t$, and hence one can shift the offset by setting $a=1$.
The slope $\beta$ can then be obtained by the ratio $\beta=-\frac{y(L_1)-y(L_2)}{x(L_1)-x(L_2)}$, where the functions $y$ and $x$ are evaluated at time $t=t_{H}^{\rm typ}$ such that the dependence on the system size $L$ enters through $t_{H}^{\rm typ}$.
Specifically,
$y(L)=\ln[P(t_{H}^{\rm typ})-P_\infty]=-\gamma \ln[c D(L)]$ and $x(L)=\ln[t_{H}^{\rm typ}(L)]$, and we express the ratios of Heisenberg times as $t_{H}^{\rm typ}(L_2)/t_{H}^{\rm typ}(L_1) = \left[ D(L_2)/D(L_1)\right]^n$.
This leads to 
\begin{equation} \label{def_beta_gamma_n}
\beta = \gamma/n\;,
\end{equation}
where $n$ is a rational positive number.
The power-law exponent $\beta$ is hence determined by the fractal dimension $\gamma$ and the scaling properties of $t_{H}^{\rm typ}$ when expressed in terms of the Hilbert-space dimension $D$.

If the scaling of $t_{H}^{\rm typ}$ with $L$ is identical to the scaling of the average $t_{H}$ with $L$, it implies $n\approx 1$, but if the spectrum exhibits level clustering or large gaps, they may lead to $n>1$.
While this derivation does not distinguish between quadratic and interacting systems, we note that by introducing $n$ in Eq.~(\ref{def_beta_gamma_n}) for interacting systems, where $D(L)$ scales exponentially with $L$, we neglect multiplicative factors that scale polynomially with $L$.
Still, as shown below, at sufficiently large $L$ these contributions can be neglected.

\begin{figure}[t!]
\centering
\includegraphics[width=\columnwidth]{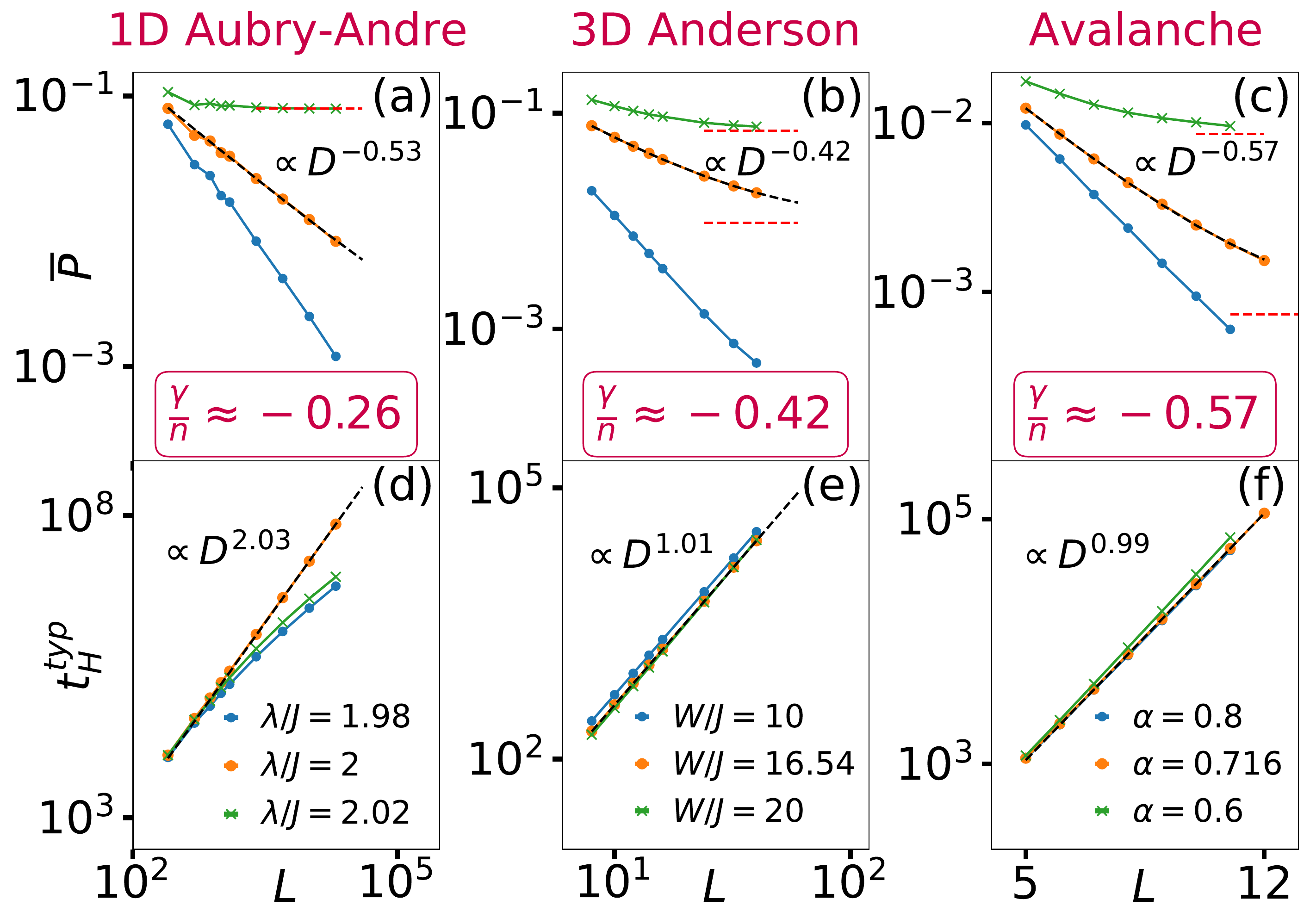}
\caption{
Scaling of $\overline{P}$ and $t_{H}^{\rm typ}$ in the models under investigation:
(a), (d) 1D Aubry-Andre model with $D = L$;
(b), (e) 3D Anderson model with $D = L^3$;
and (c), (f) avalanche model with $D = e^{(\ln 2)(N+L)}$.
Black dashed lines are fits to the data at eigenstate transitions (circles).
(a)--(c): The fractal dimension $\gamma$ is obtained using Eq.~(\ref{def_Pbar}), where the horizontal lines denote $P_\infty$.
At eigenstate transitions we get (a) $\gamma=0.53$, (b) $\gamma=0.42$, and (c) $\gamma=0.57$.
(d)--(f): The number $n$ is obtained using the ansatz $t_H^{\rm typ} \propto D^n$.
At eigenstate transitions we get (d) $n=2.03$, (e) $n=1.01$, and (f) $n=0.99$.
The ratios $\gamma/n$ given in the legends accurately match the values of $\beta$ from Fig.~\ref{figM2}, in accordance with Eq.~(\ref{def_beta_gamma_n}).
}
\label{figM3}
\end{figure}

We test predictions from Eq.~(\ref{def_beta_gamma_n}) numerically in Fig.~\ref{figM3}.
Specifically, we extract $\gamma$ and $n$ from the scaling properties of $\overline{P}$ and $t_{H}^{\rm typ}$ at eigenstate transitions in Figs.~\ref{figM3}(a)--\ref{figM3}(c) and~\ref{figM3}(d)--\ref{figM3}(f), respectively, and compare their ratios to the values of $\beta$ obtained in Figs.~\ref{figM1}(b),~\ref{figM1}(e), and~\ref{figM1}(h), finding excellent agreement.
We note that in the 1D Aubry-Andre model, the distribution of level spacings at the transition is anomalous \cite{Geisel_91}.
In Fig.~\ref{figM3}(d), we observe $t_{H}^{\rm typ} \approx D^2$, which justifies the introduction of $n\neq 1$ in Eq.~(\ref{def_beta_gamma_n}), and is consistent with $\beta \approx \gamma/2$ from Ref.~\cite{Huckestein94}.
On the other hand, in the 3D Anderson model where $\beta \approx \gamma$~\cite{Brandes96,Ohtsuki97}, see Fig.~\ref{figM3}(b), we observe $P_\infty \neq 0$ at the transition, which is a consequence of the mobility edge \cite{Schubert_05} and hence justifies the introduction of $P_\infty$ to the definition of scaled survival probability in Eq.~(\ref{eq:sur_prob_norm}).

{It is interesting to observe that $\overline{P}$ at the transition in the avalanche model [cf.~$\alpha=0.716$ in Fig.~\ref{figM3}(c)] exhibits (multi)fractal behavior, and we attribute its saturation to a small nonzero $P_\infty$ as a hallmark of the mobility edge.
In the nonergodic phase [cf.~$\alpha=0.6$ in Fig.~\ref{figM3}(c)], $\overline{P}$ saturates to a rather large value, indicating Fock space localization.
Note that the latter is a consequence of interactions and is not expected to emerge in many-body states of localized quadratic models.
}

{\it Survival probability and spectral form factor.--}
An interesting open question concerns the relation of survival probability at eigenstate transitions with the statistical properties of Hamiltonian spectra.
Recent studies of the spectral form factor (SFF) at eigenstate transitions of the 3D Anderson model~\cite{suntajs_prosen_21} and the avalanche model~\cite{suntajs_vidmar_22} observed a scale-invariant plateau in time domain that extends over several orders of magnitude.
Even though survival probability is formally not equivalent to the SFF, certain analogies can be established for the random matrices~\cite{torresherrera_garciagarcia_18, dag2022manybody} and in general~\cite{delcampo_molinavilaplana_17, dag2022manybody} (see also~\cite{SM}).
It is then reasonable to conjecture in these cases that the scale-invariant plateau in the SFF is related to the scale-invariant behavior of the survival probability.
In~\cite{SM} we numerically test this conjecture and observe that both scale-invariant phenomena occur in approximately the same time windows.

We note that the SFF in the 1D Aubry-Andre model, in contrast to the other two models, exhibits a scale-invariant {\it power-law} decay at the transition due to fractality of the eigenspectrum at the transition.  The latter emerges in nearly the same time window as a power-law decay of the survival probability~\cite{SM}.

\textit{Conclusions.--}
The new results of this Letter can be summarized in two steps.
In the first, we established scale invariance of survival probability at eigenstate transitions.
This allows us to consider scale invariance in two paradigmatic quadratic models, the 1D Aubry-Andre model and the 3D Anderson model, within the same framework.
In the second, most important step, we observe that this phenomenology also applies to ergodicity breaking transitions in interacting systems.
We note that the hallmark of the transition is scale invariance and not the mere power-law decay of the survival probability.
For quantum quenches from initial states different than those considered here, e.g., translationally invariant plane waves, power-law decay may not be present, however, signatures of scale invariance may still emerge~\cite{SM, mh_lv_inprep}.

{The main advantage of introducing scale invariance at eigenstate transitions is to establish a tool to detect the transition point in time domain at relatively short times.
These times are much shorter than the characteristic relaxation time (also denoted as the Thouless time), which in the interacting models scales exponentially with $L$ at the transition.
This opens new possibilities to characterize and detect ergodicity breaking phenomena, in particular, to extend our framework to few-body observables measured in experiments.
}

We acknowledge discussions with P. Prelovšek and T. Prosen and J. Šuntajs.
We acknowledge the support of the Slovenian Research Agency (ARRS), Research Core Fundings Grants P1-0044 and N1-0273.
We gratefully acknowledge the High
Performance Computing Research Infrastructure Eastern
Region (HCP RIVR) consortium~\cite{vega1} and
European High Performance Computing Joint Undertaking (EuroHPC JU)~\cite{vega2}  for funding
this research by providing computing resources of the
HPC system Vega at the Institute of Information sciences
~\cite{vega3}.

\bibliographystyle{biblev1}
\bibliography{references}


\newpage
\phantom{a}
\newpage
\setcounter{figure}{0}
\setcounter{equation}{0}
\setcounter{table}{0}

\renewcommand{\thetable}{S\arabic{table}}
\renewcommand{\thefigure}{S\arabic{figure}}
\renewcommand{\theequation}{S\arabic{equation}}
\renewcommand{\thepage}{S\arabic{page}}

\renewcommand{\thesection}{S\arabic{section}}

\onecolumngrid

\begin{center}

{\large \bf Supplemental Material: \\
Scale-invariant survival probability at eigenstate transitions}\\

\vspace{0.3cm}

\setcounter{page}{1}

Miroslav Hopjan$^{1}$ and Lev Vidmar$^{1,2}$\\
$^1${\it Department of Theoretical Physics, J. Stefan Institute, SI-1000 Ljubljana, Slovenia} \\
$^2${\it Department of Physics, Faculty of Mathematics and Physics, University of Ljubljana, SI-1000 Ljubljana, Slovenia}

\end{center}

\vspace{0.6cm}

\twocolumngrid

\label{pagesupp}

\section{Transition in the avalanche model}

\begin{figure}[!b]
\centering
\includegraphics[width=\columnwidth]{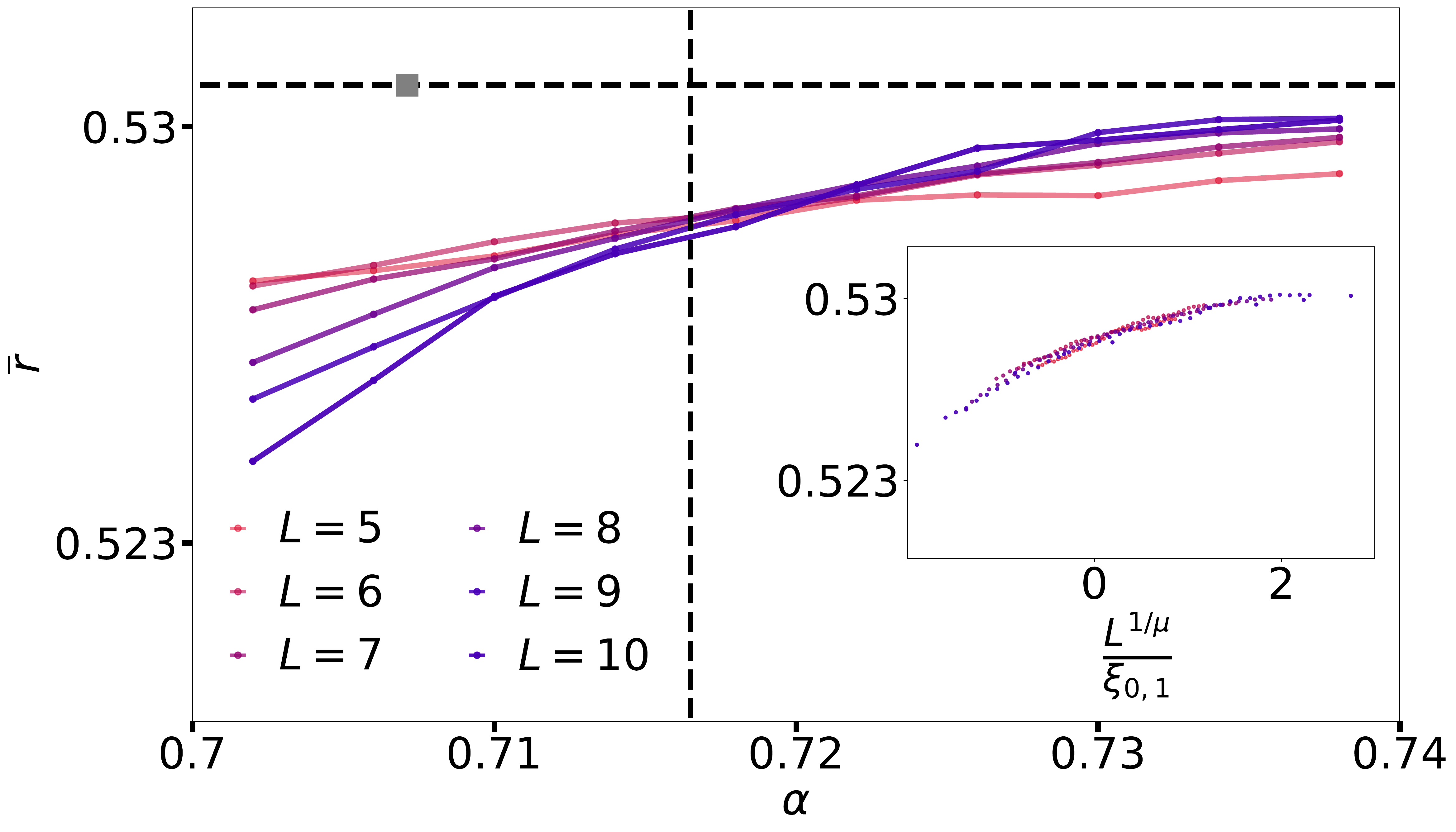}
\caption{Results for the mean level spacing ratio $\overline{r}$ from Eq.~(\ref{gap_ratio}), as a function of $\alpha$ at different $L$. 
Inset: The scaling collapse
of $\bar r$ versus $L^{1/\mu}/\xi_{0,1}$, see the main text for details.
The gray square in the main panel denotes the analytical prediction for the transition point $\bar\alpha=1/\sqrt{2}$, while the vertical dashed line is the numerically extracted transition point $\alpha_c = 0.716$ obtained from the scaling collapse in the inset.
}
\label{figS0}
\end{figure}

While the values of the transition points in the 1D Aubry-Andre model and the 3D Anderson model are known to high accuracy, we here establish the transition point in the avalanche model from Eq.~(\ref{eq:hamQSM}) that is used in the analysis in Figs.~\ref{figM1}(h) and~\ref{figM2} in the main text.

As an indicator of the transition, we use the ratio of consecutive level spacings of the many-body spectrum,
\begin{equation} \label{gap_ratio}
    r_\nu=\frac{\min\{\delta E_{\nu+1} ,\delta E_\nu\}}{\max\{\delta E_{\nu+1} ,\delta E_\nu\}}
\end{equation}
where $\delta E_{\nu} = E_{\nu+1} -E_\nu$ is the nearest level spacing. We define the mean ratio as $\bar r = \langle\langle r_\nu \rangle_{\nu} \rangle_{H}$, with $\langle ... \rangle_\nu$ denoting the average over pairs of spacings of nearest levels and $\langle ... \rangle_H$ denoting the average over Hamiltonian realizations. For each Hamiltonian realization we average over 500 energy states close to the mid-spectrum, and we then average over 10000 Hamiltonian realizations.

Results for $\bar r$ are shown in Fig.~\ref{figS0}.
The main panel shows $\bar r$ versus $\alpha$, while the inset shows the scaling collapse of $\bar r$ versus $L^{1/\mu}/\xi_{0,1}$, where  $\xi_0=1/\ln(\alpha/\alpha_c)^{2}$ for $\alpha>\alpha_c$ and $\xi_1=1/\ln(\alpha_c/\alpha)^{2}$ for $\alpha<\alpha_c$~\cite{suntajs_vidmar_22}.
Using the cost function minimization procedure introduced in~\cite{suntajs_bonca_20b}, we obtain the optimal fitting parameters $\mu\approx 0.6 $ and $\alpha_c\approx 0.716$. 
The numerical estimate for the transition point $\alpha_c\approx 0.716$ is very close (within 2\%) to the analytical estimate $\bar\alpha=1/\sqrt{2}\approx 0.707$ using the hybridization condition from Ref.~\cite{deroeck_huveneers_17}. 

On a technical side, we note that in the definition of the avalanche model in Eq.~(\ref{eq:hamQSM}) we use $N=5$ and we set the coupling to the closest spin outside the dot to one (since $u_0=0$).
This definition is different from a recent study in Ref.~\cite{suntajs_vidmar_22}, which used $N=3$ and the transition point estimate was $\alpha_c\approx0.75$. 
We observe that by increasing $N$ in finite-size calculations, the values of the transition point are quantitatively closer to the analytical prediction $\bar\alpha=1/\sqrt{2}$.

\section{Connection to the Spectral form factor.}

We next compare the behaviour of the survival probability $p_{}^{}(\tau)$ shown in the main text  to the behaviour of the spectral form factor (SFF).
Recently, the characterisation of the eigenstate transitions using the SFF has been carried out
for the 3D Anderson model~\cite{suntajs_prosen_21} and the avalanche model~\cite{suntajs_vidmar_22}. We note that, therein, the SFF was 
defined using unfolded spectra and a Gaussian filtering function. In both studies a broad plateau of the SFF at the transition was found and therefore the SFF universality at the transition has been conjectured~\cite{suntajs_vidmar_22}. 

\begin{figure}[t!]
\centering
\includegraphics[width=\columnwidth]{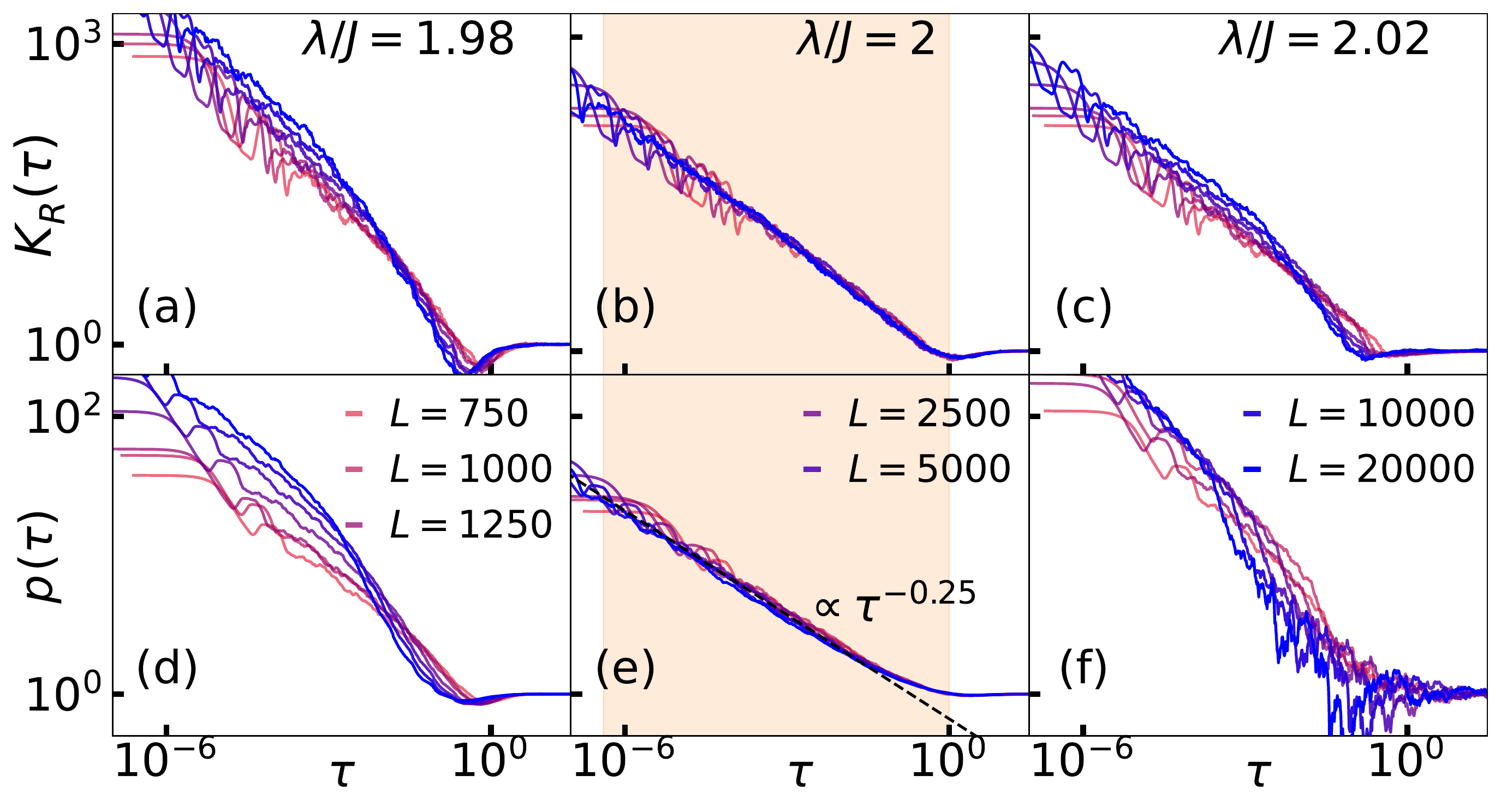}
\caption{
(a)-(c) The raw SFF $K_{R}^{}(\tau)$ from Eq.~(\ref{eq:sff_raw}) as function of the scaled time $\tau=t/t_{\rm H}^{\rm typ}$ in the 1D Aubry-Andree model at different system sizes ($D=L$), plotted (a) in the delocalized regime, (b) at the transition point, and (c) in the localized regime.
(d)-(f) The corresponding survival probability $p(\tau)$ [the same results as in Figs.~\ref{figM1}(a)-\ref{figM1}(c) of the main text].
The shaded areas in (b) and (e) denote the scale invariance in $p_{}^{}(\tau)$ for the largest system size $L=20000$.}
\label{figS1}
\end{figure}

Here we consider the raw SFF computed from the raw Hamiltonian eigenvalues,
\begin{equation}
\label{eq:sff_raw}
K_{R}^{}(t)=\frac{1}{D}\bigl< \big| \sum_{\nu=1}^D  e^{-iE_\nu t}  \big|^2 \bigl>_H,
\end{equation}
where no spectral unfolding and filtering function are applied.
We show that, still, the SFF defined in this way retains the universality observed in Refs.~\cite{suntajs_prosen_21, suntajs_vidmar_22}.
The reason to use the raw SFF $K_{R}^{}(t)$ is that it naturally connects to the 
survival probability $p_{}^{}(t)$ in Eq.~\eqref{eq:sur_prob_norm}.
Indeed, $K_{R}^{}(t)$ can be interpreted as $p(t)$ where each initial state 
$|m \rangle$ in Eq.~\eqref{eq:sur_prob} is replaced by the infinite-temperature pure state $ |m^{}_{T=\infty} \rangle =\sum_\nu D^{-1/2} |\nu \rangle $.
Then, the values of long-time limits in Eq.~\eqref{eq:sur_prob_norm} are $\overline{P^{}}=1/D$ and $P_{\infty}=0$, and $p_{}^{}(t)$ hence
reduces to $K_{R}^{}(t)$. 
To discuss the connection of $K_{R}^{}(t)$ to the scale invariance of $p_{}^{}(\tau)$, we plot $K_{R}^{}$ as a function of the scaled time $\tau = t/t_{\rm H}^{\rm typ}$.

\begin{figure}[!b]
\centering
\includegraphics[width=\columnwidth]{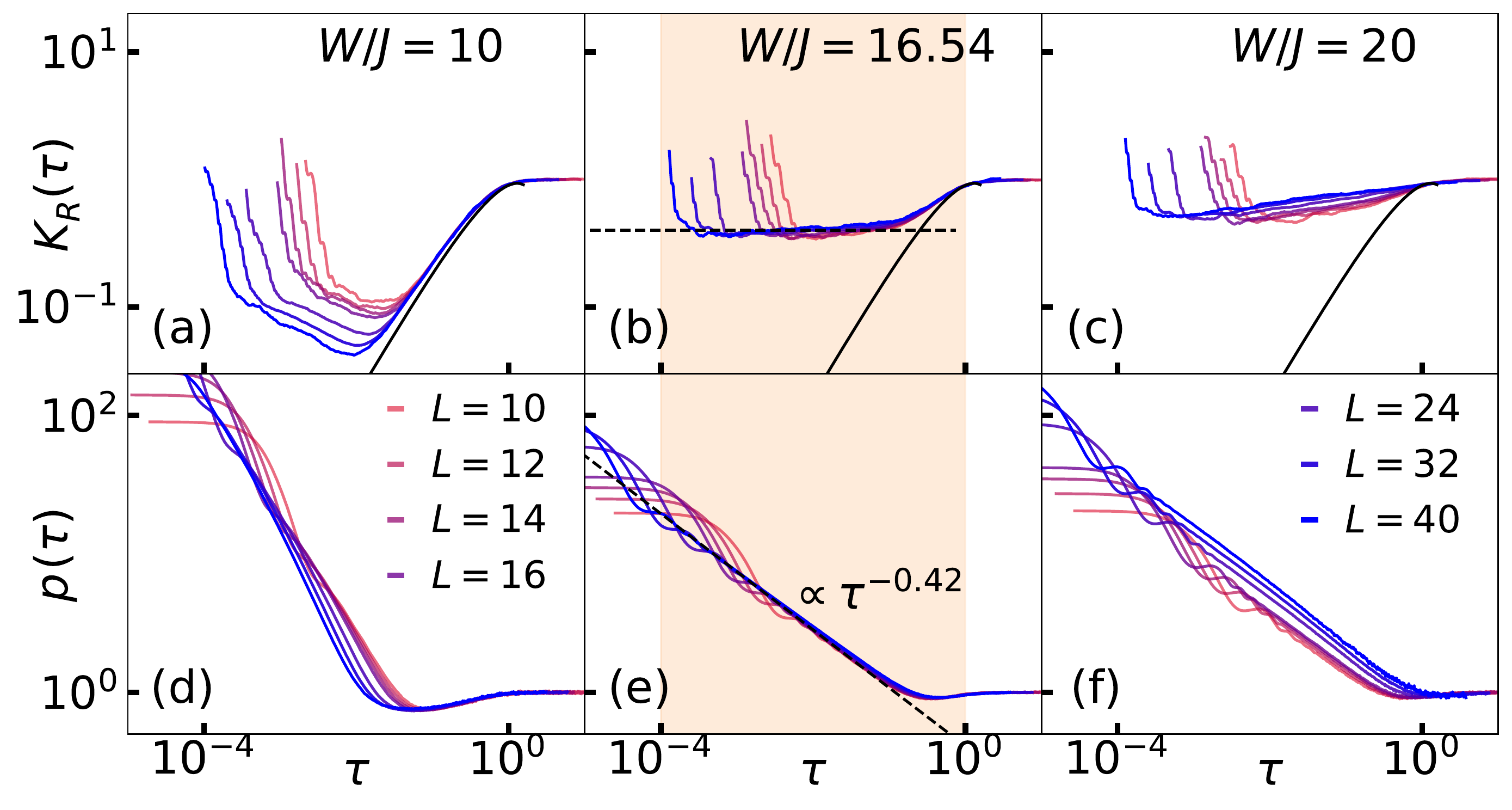}
\caption{
(a)-(c) The raw SFF $K_{R}^{}(\tau)$ from Eq.~(\ref{eq:sff_raw}) as function of the scaled time $\tau=t/t_{\rm H}^{\rm typ}$ in the 3D Anderson model at different system sizes ($D=L^3$), plotted (a) in the delocalized regime, (b) at the transition point, and (c) in the localized regime.
The solid line in (b) denotes the GOE result $K_{\rm GOE}^{}(\tau)=2\tau-\tau \ln(1-2\tau)$, while the dashed line in (b) denotes a plateau of $K_{R}^{}(\tau)$. 
(d)-(f) The corresponding survival probability $p(\tau)$ [the same results as in Figs.~\ref{figM1}(d)-\ref{figM1}(f) of the main text].
The shaded areas in (b) and (e) denote the time interval of the scale invariance in $p_{}^{}(\tau)$  for the largest system size $L=40$.}
\label{figS3}
\end{figure}

In Figs.~\ref{figS1}(a)-\ref{figS1}(c) we plot the raw SFF $K_{R}^{}(\tau)$ for the 1D Aubry-Andre model and compare it to the survival probability $p_{}^{}(\tau)$, see Figs.~\ref{figS1}(d)-\ref{figS1}(f),
which contain the same results as in Figs.~\ref{figM1}(a)-\ref{figM1}(c) of the main text.
We observe that at the transition point, see Figs.~\ref{figS1}(b) and~\ref{figS1}(e), the scale invariant power-law behavior in $p_{}^{}(\tau)$ is accompanied by a similar behavior in $K_{R}^{}(\tau)$.
Away from the transition, see Figs.~\ref{figS1}(a) and~\ref{figS1}(c), such scale invariance is lost.
We note that the power-law behavior in $K_R(\tau)$ in Fig.~\ref{figS1}(b) in the 1D Aubry-Andre model is different from the behavior in the 3D Anderson model and the avalanche model, which we discuss below. 

\begin{figure}[t!]
\centering
\includegraphics[width=\columnwidth]{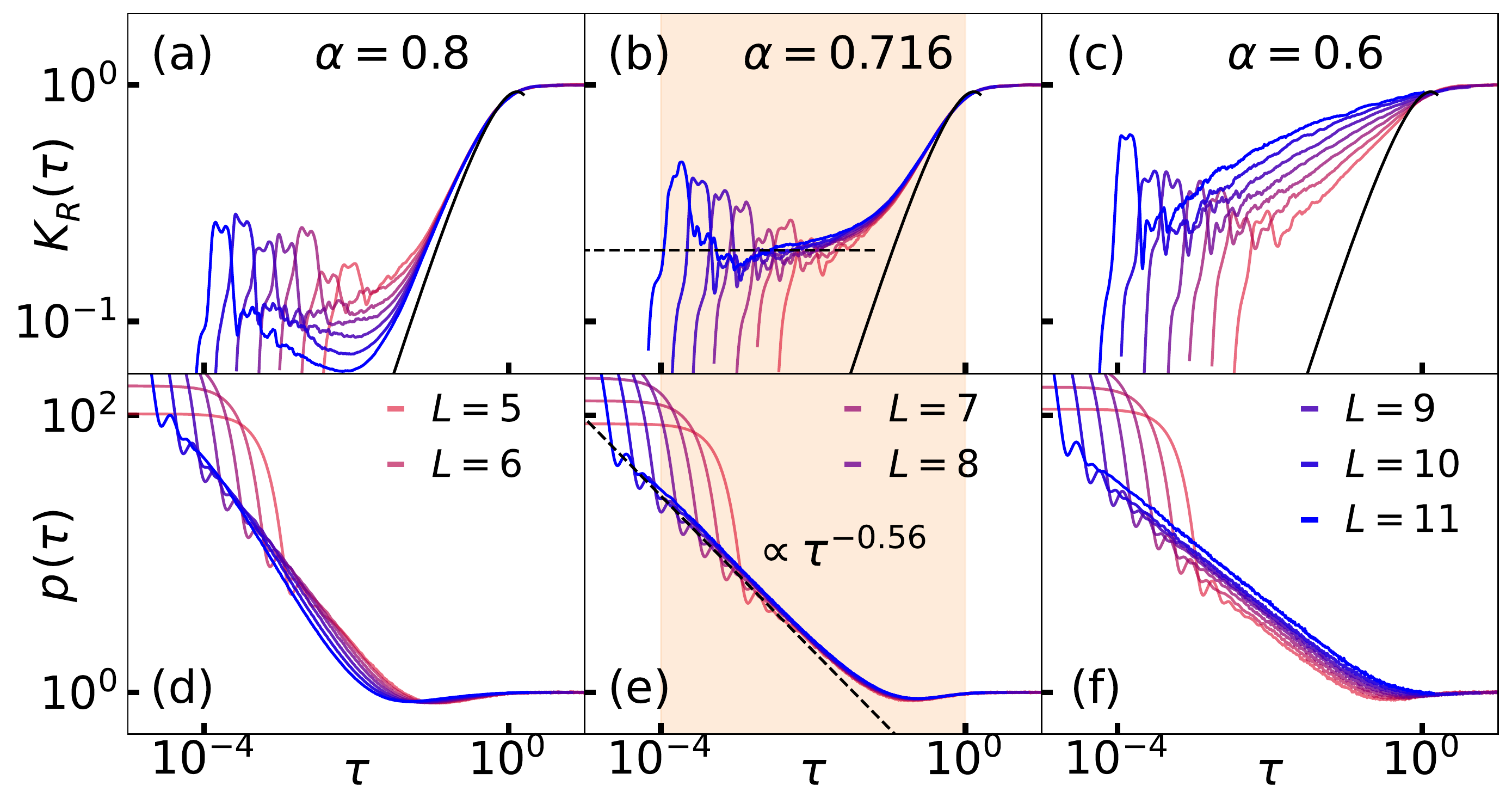}
\caption{The raw SFF $K_{R}^{}(\tau)$ from Eq.~(\ref{eq:sff_raw}) as function of the scaled time $\tau=t/t_{\rm H}^{\rm typ}$ in the avalanche at different system sizes ($D=2^{N+L}$), plotted (a) in the delocalized phase, (b) at the transition point, 
and (c) in the localized phase.  The black lines denotes the GOE result $K_{\rm GOE}^{}(\tau)=2\tau-\tau \ln(1-2\tau)$. The dashed line in (b) denotes a plateau of $K_{R}^{}(\tau)$. 
(d)-(f) The corresponding survival probability $p(\tau)$ [the same results as in Figs.~\ref{figM1}(g)-\ref{figM1}(i) of the main text].
The shaded areas in (b) and (e) denote the time interval of the scale invariance in $p_{}^{}(\tau)$  for the largest system size $L=12$.}
\label{figS4}
\end{figure}

In Fig.~\ref{figS3}, we  compare the raw SFF $K_{R}^{}(\tau)$ and $p_{}^{}(\tau)$ in the 3D Anderson model. The results for $p_{}^{}(\tau)$ are the same as in
Figs.~\ref{figM1}(d)-\ref{figM1}(f) in the main text.
The behaviour of the SFF was discussed in Ref.~\cite{suntajs_prosen_21}.
This is, below the transition, the SFF follows the 
GOE prediction for times larger than the Thouless time.
More importantly, at the transition, the SFF exhibits a broad plateau before it reaches the Thouless time.
This is also observed here for the raw SFF, see Figs.~\ref{figS3}(a)-\ref{figS3}(b).
(Note that below the transition,  the raw SFF $K_{R}^{}(\tau)$ approaches the GOE prediction with a small deviation, the latter being an artefact of using the raw spectrum.)
Remarkably, at the transition the time window of the broad scale invariant plateau in $K_{R}^{}(\tau)$, see Fig.~\ref{figS3}(b), almost exactly corresponds to the time window of the scale invariant power-law decay of $p_{}^{}(\tau)$ observed in Fig.~\ref{figS3}(e).

Finally, in Fig.~\ref{figS4}, we  compare the raw SFF $K_{R}^{}(\tau)$ with $p_{}^{}(\tau)$ for the avalanche model. The results for $p_{}^{}(\tau)$ are the same as in
Figs.~\ref{figM1}(g)-\ref{figM1}(i) in the main text.
When compared to the 3D Anderson model, the main features are qualitatively similar.
For the avalanche model, the behaviour of the SFF was discussed in Ref.~\cite{suntajs_vidmar_22}.
Also for this model, in the ergodic phase, the SFF follows the GOE prediction for times larger than the Thouless time, and at the transition the scale invariant plateau emerges.
The same features are visible in the raw SFF $K_{R}^{}(\tau)$, see Figs.~\ref{figS4}(a)-\ref{figS4}(c).
Interestingly, for quenches from fully polarized spins in the avalanche model, the time window of the scale invariant plateau in $K_{R}^{}(\tau)$, see Fig.~\ref{figS4}(b), is actually shorter than the time window of the scale invariant power-law decay of $p_{}^{}(\tau)$ observed in Fig.~\ref{figS4}(e). This suggest that the time range of scale invariance may, at least for the system sizes under investigation, differ for different initial states.

\section{Quenches from initial plane waves}

\begin{figure}[!b]
\centering
\includegraphics[width=\columnwidth]{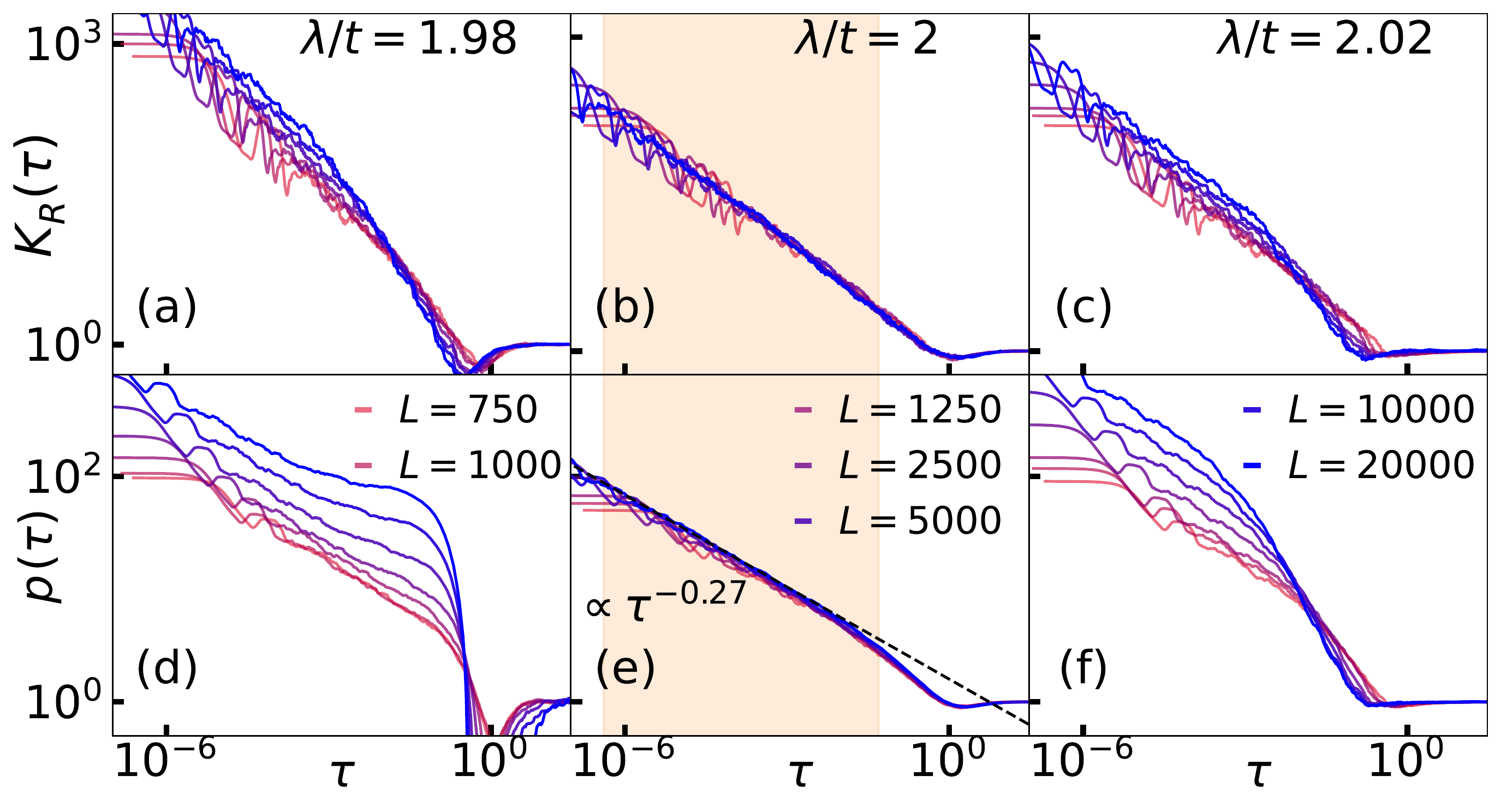}
\caption{The raw SFF $K_{R}^{}(\tau)$ from Eq.~(\ref{eq:sff_raw}) as function of the scaled time $\tau=t/t_{\rm H}^{\rm typ}$ in the 1D Aubry-Andre model at different system sizes ($D=L$), plotted (a) in the delocalized regime, (b) at the transition point, and (c) in the localized regime.
(d)-(f) The corresponding survival probability $p(\tau)$, for which the initial states are single-particle plane waves, as described in the text.
The dashed line in (e) denotes the power-law fit $p(\tau) = a \tau_{}^{-\beta}$ to the results at the transition, and we obtain $\beta=0.27$.
The shaded areas in (b) and (e) denote scale invariance in $p_{}^{}(\tau)$ for the largest system size $L=20000$.}
\label{figS5}
\end{figure}

In the main text, we discuss the behaviour of the survival probability for quenches from localized states (in particular, localized states in real space for the 1D Aubry-Andre and 3D Anderson model), and show its scale-invariant properties at the transition.  Here we discuss an opposite limit, in which the initial states are fully delocalized plane-wave states $ |k^{}_{} \rangle$, i.e. states constructed as $ |k^{}_{} \rangle = \frac{1}{\sqrt{D}}\sum_m e^{-ikm} |m \rangle $ where $|m \rangle$ are the initially localized states considered above and $k=0,\dots,D-1$.

\begin{figure}[!t]
\centering
\includegraphics[width=\columnwidth]{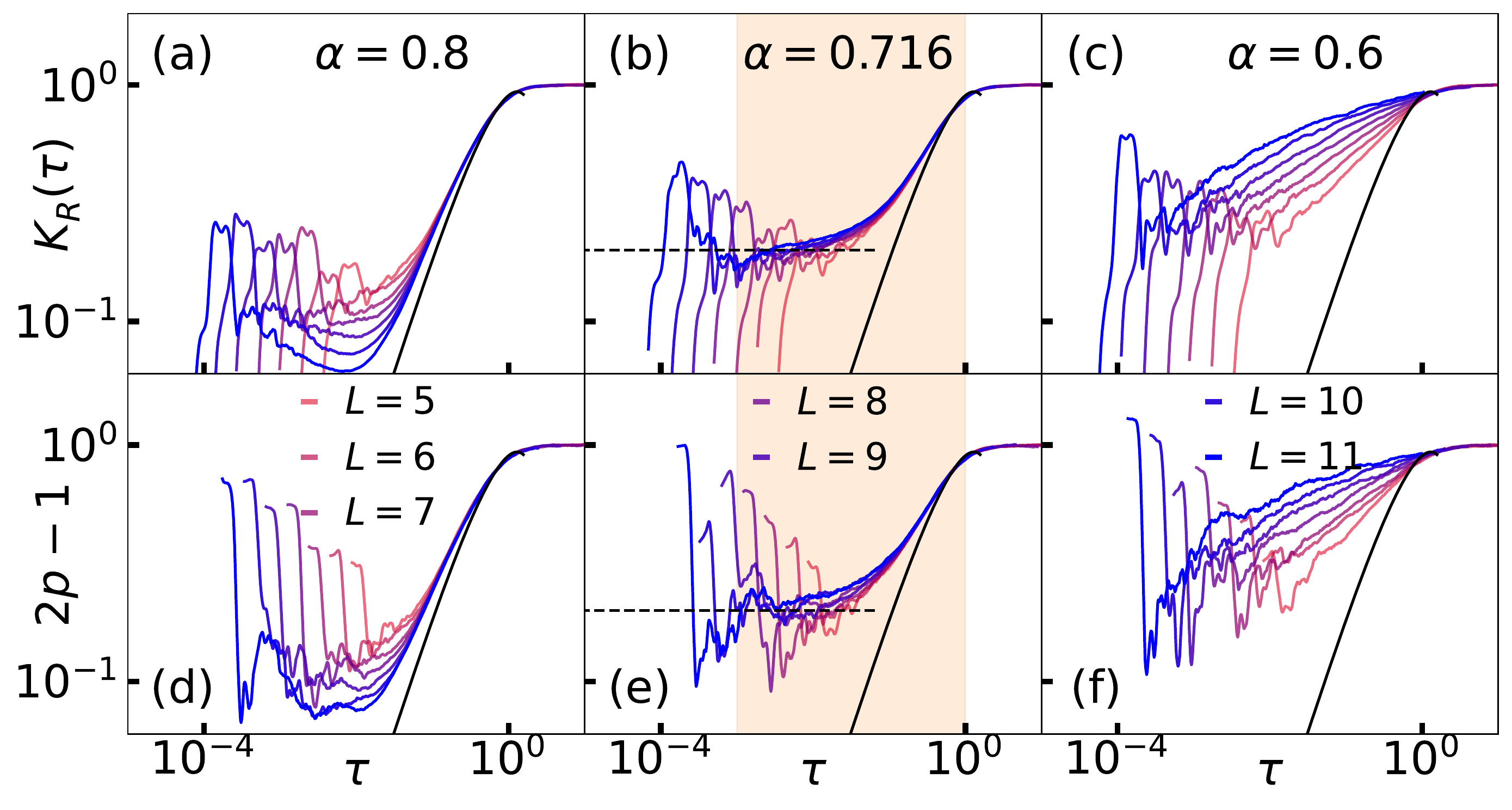}
\caption{The raw SFF $K_{R}^{}(\tau)$ from Eq.~(\ref{eq:sff_raw}) as function of the scaled time $\tau=t/t_{\rm H}^{\rm typ}$ in the avalanche model at different system sizes ($D=2^{N+L}$), plotted (a) in the delocalized phase, (b) at the transition point, and (c) in the localized phase.
(d)-(f) The corresponding survival probability $2p(\tau)-1$, for which the initial states are plane waves, as described in the text.
The shaded areas in (b) and (e) denote the scale invariance in $p_{}^{}(\tau)$  for the largest system size $L=11$.}
\label{figS6}
\end{figure}

We first discuss the 1D Aubry-Andre model. In Fig.~\ref{figS5}, we show $p_{}^{}(\tau)$ for initial conditions that are delocalized plane waves, and we compare it to the raw SFF $K_{R}^{}(\tau)$. In spite of the change of initial conditions, one may draw similar conclusions about the scale invariance of $K_{R}^{}(\tau)$ and $p_{}^{}(\tau)$ at the transition as in the case  of initial localized states, see Fig.~\ref{figS1}. The scale invariance takes place in a nearly identical time interval as in Fig.~\ref{figS1}.
Moreover, the exponent $\beta$ of the power-law decay is very similar in both cases: we get $\beta=0.25$ in Fig.~\ref{figS1} and $\beta=0.27$ in Fig.~\ref{figS5}.
In the case of the 1D Aubry-Andre model, the eigenstate transition occurs between eigenstates that are localized in quasimomentum space and eigenstates that are localized in real space~\cite{Aubry80, Suslov82}. The transition point is a self-dual point, suggesting that both initial conditions, the fully localized and fully delocalized ones, may have similar overlaps with eigenstates at the transition. This may explain the similarity of the power-law decays from both initial conditions.

Next, we discuss the quenches from plane waves for the second type of transition, namely, the transition that occurs between the eigenstates whose properties are consistent with predictions of the GOE, and the localized eigenstates that are accompanied with Poisson statistics of eigenenergies. We here show results for the avalanche model, while the 3D Anderson exhibits qualitatively similar results.
In Fig.~\ref{figS6}, we compare $p_{}^{}(\tau)$ to the raw SFF $K_{R}^{}(\tau)$. 
In particular, we show $2p_{}^{}(\tau)-1$ instead of $p_{}^{}(\tau)$, which is inspired by the exact relation $K_{R}^{}(\tau)=2p_{}^{}(\tau)-1$ 
for the GOE matrices~\cite{torresherrera_garciagarcia_18, dag2022manybody}. 
Indeed, we observe that in the ergodic regime, see  Figs.~\ref{figS6}(a) and~\ref{figS6}(d), the behavior of the SFF $K_{R}^{}(\tau)$ is similar to the behavior of $2p_{}^{}(\tau)-1$.
Surprisingly, this similarity is also observed at the transition point, see  Figs.~\ref{figS6}(b) and~\ref{figS6}(e).
In the latter case, the scale invariant plateaux are building up in both quantities $K_{R}^{}(\tau)$  and
$2p_{}^{}(\tau)-1$ at roughly the same values [see the horizontal dashed lines in Figs.~\ref{figS6}(b) and~\ref{figS6}(e)], and in the same time interval.
This result demonstrates that the scale invariance of $p_{}^{}(\tau)$ may emerge without the presence of the power-law decay.
In the nonergodic phase ($\alpha<\alpha_c$) the qualitative similarity between  $K_{R}^{}(\tau)$ and $2p_{}^{}(\tau)-1$ persist, see Figs.~\ref{figS6}(c) and~\ref{figS6}(f).

\section{Details of averaging}

For the data shown in Figs.~\ref{figM1} and~\ref{figM3} of the main text and in Figs.~\ref{figS1}-\ref{figS5} of this Supplemental material we average over $500$ Hamiltonian realizations for all models and all system sizes, except for $L=12$ of the avalanche model.
In the latter case, see Fig.~\ref{figM2}, the size of the Hamiltonian matrix is $D=2^{L+N}=131072$ and a satisfactory convergence is obtained by averaging over $15$ Hamiltonian realizations ($\approx 10^6$ initial states).
For the spectral form factor, we additionally use running averages to reduce time fluctuations around the mean value.

\newpage

\end{document}